\documentclass[preprintnumbers,superscriptaddress,amsmath,amssymb,prf]{revtex4-2}
\usepackage{graphicx}
\usepackage{dcolumn}
\usepackage{soul}
\usepackage[usenames,dvipsnames]{xcolor}
\usepackage{mathptmx}
\usepackage{amssymb}
\usepackage{array}
\usepackage{amsmath}
\usepackage{setspace}
\usepackage{hyperref}
\usepackage{comment}
\usepackage{lineno}
\usepackage[italicdiff]{physics}

\begin{document}
\preprint{ \color{Blue} \today}

\title{\Large \color{Blue} Stretching water between two grooves}

\author{M.Leonard}
\affiliation{GRASP, Institute of Physics B5a, University of Li\`ege, 4000 Li\`ege, Belgium}

\author{D.Maity}
\affiliation{King Abdullah University of Science and Technology, Physical Sciences and Engineering, Splash Lab, Saudi Arabia}

\author{N.Vandewalle}
\affiliation{GRASP, Institute of Physics B5a, University of Li\`ege, 4000 Li\`ege, Belgium}

\author{T.Truscott}
\affiliation{King Abdullah University of Science and Technology, Physical Sciences and Engineering, Splash Lab, Saudi Arabia}

%%%%%%%%%%%%%%%%%%%%%%%%%%%%%  abstract Before \maketitle

\begin{abstract}

Controlling water motion on surfaces is critical for applications ranging from thermal management, passive water harvesting, to self-cleaning coatings. Yet stabilising continuous water films, desirable for their high surface coverage and drainage capacity, remains challenging with pure water, due to its high surface tension.
Existing strategies rely on extreme wettability achieved by coating or fine-scale patterning, which are costly, fragile, or complex to scale. A robust, purely geometric solution is still lacking.
We demonstrate that a pair of laser-engraved grooves on a moderately hydrophilic vertical substrate can laterally anchor water and stretch a continuous thin film, without any other surface treatment.
Once stabilised, the film extends vertically over 100 capillary lengths ($> 30\rm{cm}$), with thickness tunable via both flow rate and groove geometry. At the groove extremities, the end of anchoring triggers a cyclic instability, characterised by film rupture, retraction, and droplet release. The thickness of the film and retraction height obey predictive models, while droplet mass varies systematically with spacing and surface tension.
This groove-based method offers a straightforward and scalable approach to creating, sustaining, and controlling thin water films. It opens new directions for passive liquid control in condensation, surface cleaning, and 2D millifluidic systems.

\end{abstract}

\maketitle

\section{Introduction}

Controlling how water moves across surfaces is essential in both natural and engineered systems. In arid regions, passive water collection devices aim to harvest moisture from the atmosphere through fog \cite{abdul-wahab_feasibility_2007,estrela_prospective_2009, henschel_ecophysiology_2008, ritter_fog_2008,bintein_kirigami_2023} and dew collection \cite{beysens_using_2003,lekouch_rooftop_2012,haechler_exploiting_2021,jacobs_passive_2008,maestre-valero_comparative_2011,muselli_dew_2009}. In industry, the efficient removal of condensate is critical for thermal management in systems such as heat exchangers, where stagnant water can degrade performance \cite{hu_review_2021, orejon_simultaneous_2017}. Even standard coatings, such as paint or water-repellent finishes, are designed to control how water behaves, whether it spreads out or forms droplets, to keep surfaces clean and working properly \cite{zhu_universal_2021, drelich_hydrophilic_2011, geyer_when_2020, choi_superhydrophilic_2017}. Across these diverse applications, the challenge remains the same: how to guide water where we want it to go, and when we want it to move, using surfaces that are both robust and scalable.

On smooth vertical surfaces, water typically adopts one of two classical morphologies depending on the flow rate: isolated droplets at low flow, and continuous rivulets at higher flow \cite{podgorski_corners_2001, snoeijer_cornered_2007, ghezzehei_constraints_2004}. In contrast, many natural surfaces exhibit complex behaviours thanks to surface texture. Plants and animals use hairs \cite{chen_ultrafast_2018, duprat_wetting_2012}, spines \cite{andrews_three-dimensional_2011, bai_cactus_2020,lee_multiple_2022}, and ridges \cite{ang_enhancing_2019,chen_continuous_2016} to manipulate water, enabling directional transport \cite{kern_twisted_2024, van_hulle_droplet_2024,khattak_directed_2024}, enhanced retention \cite{pan_drop_2018, leonard_grooves_2025}, or self-cleaning \cite{barthlott_purity_1997}. Among these, grooves stand out for their ability to pin contact lines \cite{chen_anisotropy_2005, dai_hydrophilic_2018}, guide liquid in well-defined pathways \cite{comanns_directional_2015, dhiman_self-transport_2018,van_hulle_effect_2023} and promote capillary imbibition \cite{leonard_grooves_2025,bintein_grooves_2019, seemann_wetting_2005}. At smaller scales, nano- and microstructured surfaces can promote rapid liquid spreading \cite{nakajima_superhydrophilicity_2016, zhang_fabrication_2014}, enabling the formation of thin water films \cite{poualvarez_efficient_2025,dorbolo_wetting_2021}, a distinct state that lies between discrete droplets and continuous rivulets.

Thin films offer several attractive properties: they maximise surface/volume ratio, provide continuous and stable drainage paths, and enhance heat and mass transfer. Yet, despite these advantages, thin films remain difficult to form and sustain, particularly with pure water, which is characterised by a strong surface tension $\sigma$. Achieving a stable film typically requires extreme surface wettability, often obtained through chemical coatings or nano- and micro-patterning \cite{quere_wetting_2008, shibuichi_super_1996}. Such structures, even if ubiquitous in nature \cite{koch_multifunctional_2009}, remain difficult and costly to reproduce synthetically. Moreover, these treatments should be effective across the entire surface, as even minor defects may compromise film stability \cite{couvreur_role_2012,vrij_rupture_1968}. Most treatments are short-lived, environmentally unsustainable, or prohibitively expensive to scale.

In this work, we introduce a novel method for creating thin films. Imagine holding a stretched elastic sheet between your hands. The moment you let go, it snaps back to its rest state. A thin film of water on a smooth surface behaves similarly; it retracts and collapses. To keep the sheet stretched, you only need to fix its edges. Can the same idea apply to water? Yes, and to do so, we will use grooves.

We show that this geometric feature, a pair of laser-engraved vertical grooves spaced a few millimetres apart, is sufficient to anchor and stabilise a thin film of water on a moderately hydrophilic substrate (PMMA), as illustrated in Figure \ref{fig: setup} (a). Unlike previous strategies that require treating the entire surface, our method uses localised geometry to maintain a continuous film that extends over heights exceeding 100 capillary lengths. The resulting film is not only robust but also tunable, as its thickness can be continuously adjusted by varying the flow rate or the groove geometry. At the bottom of the grooved region, where lateral confinement abruptly ends, the film becomes unstable. This transition triggers a controlled, localised rupture and the release of a droplet, initiating periodic dripping governed by groove spacing.

These findings demonstrate that geometric anchoring can hold a thin water film in place. With only a pair of grooves, we can define the film’s shape, tune its thickness, and trigger regular droplet release through localised instability. This approach provides a simple and scalable alternative to superhydrophilic materials, enabling robust and passive water control in systems ranging from condensation management to surface cleaning.

\section{Experimental Setup}

\begin{figure}
    \centering
    \includegraphics[width=1\linewidth]{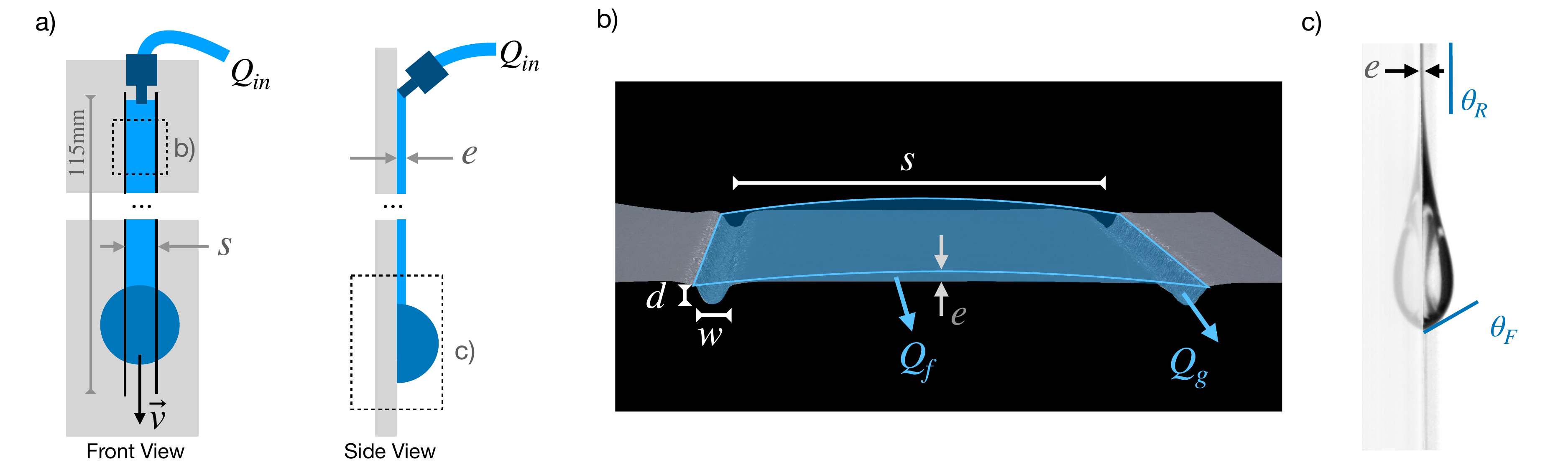}
    \caption{Experimental setup and groove geometry.
(a) Front and side views of the acrylic substrate (TroGlass Clear, $3\,\mathrm{mm}$ thick) engraved with two vertical grooves spaced by a distance $s$. The grooves, $115\,\mathrm{mm}$ long, terminate a few centimetres above the bottom edge to trigger film rupture. A syringe pump delivers water between the grooves through a $0.8\,\mathrm{mm}$ nozzle at a flow rate $Q_{\mathrm{in}}$. The side view illustrates how the thin film of thickness $e$ connects the syringe tip to the droplet near the groove termination.
(b) Groove geometry visualised using a Keyence VHX optical microscope. The three defining parameters are groove depth $d$, width $w$, and centre-to-centre spacing $s$. The example shown corresponds to a substrate with $s = 1.75\,\mathrm{mm}$ and $d/w = 0.54$. Arrows illustrate the partitioning of the injected flow $Q_{\mathrm{in}}$ into a film contribution $Q_{\mathrm{f}}$ and groove contribution $Q_{\mathrm{g}}$.
(c) Side-view image showing a retracting droplet at the lower end of the film. We measure the front contact angle $\theta_{\mathrm{F}}$ throughout the periodic instability. In contrast, we assume the rear angle to be zero, given the continuous and smooth connection with the film.}
    \label{fig: setup}
\end{figure}

To study the properties and the edge instability of vertical water films anchored by grooves, we designed a minimal but robust experimental setup. The substrate is a transparent acrylic plate (TroGlass Clear, $3\,\mathrm{mm}$ thick, colored in light grey in Figure \ref{fig: setup} (a)) engraved with two vertical grooves using a $\rm{CO}_2$ laser (Trotec Speedy 100). Each groove extends over $115\,\mathrm{mm}$ and terminates several centimetres above the lower edge to trigger film rupture, as represented in Figure \ref{fig: setup} (a). The plates are moderately hydrophilic, with receding and advancing contact angles of $\theta_R = 48\pm4^\circ$ and $\theta_A = 78\pm5^\circ$, respectively, measured using the droplet inflation/deflation method on horizontal plates.

We define the groove geometry by three key parameters: the groove depth $d$, width $w$, and centre-to-centre spacing $s$, represented in Figure \ref{fig: setup} (b). We studied three groove geometries with physical dimensions $(d, w) = (0.08, 0.15)$, $(0.21, 0.19)$, and $(0.38, 0.21)\,\mathrm{mm}$, corresponding to shallow ($d/w = 0.54$), intermediate ($1.13$), and deep ($1.80$) aspect ratios, respectively. The groove spacing $s$ was varied independently from $0.75$ to $2.25\,\mathrm{mm}$. These dimensions were measured with a Keyence VHX optical microscope.

Water was injected between the grooves through a $0.8\,\mathrm{mm}$ diameter nozzle connected to a syringe pump, with flow rates $Q_{\mathrm{in}}$ ranging from $0.33$ to $10\,\mathrm{mm^3/s}$. To ensure continuous liquid connection from the syringe tip to the groove ends, grooves were prewetted before each trial by sliding droplets along their length, either manually or by gravity. In some experiments, $0.01\,\mathrm{g}$ of SDS was added per litre of water to reduce surface tension from $\sigma_{\rm{water}} = 72\,\mathrm{mN/m}$ to $\sigma_{\rm{SDS}} = 63\,\mathrm{mN/m}$. Further lowering the surface tension suppresses film retraction after breakup, eliminating the dynamics that are central to this study.

To estimate film thickness, we used particle tracking velocimetry (PTV). $20\,\mu\mathrm{m}$ tracer particles were suspended in the water and imaged using a high-resolution charged coupled device (CCD) camera at $113\,\mathrm{fps}$. The field of view spanned a $25\,\mathrm{mm}$ segment centred within the grooved region. Velocity fields were extracted by subdividing the film into rectangular bins and averaging particle velocities within each region.

To track droplet formation and film retraction, we used a second CCD camera placed behind the substrate, with front lighting to enhance the droplet edges. Images were acquired at $13.6\,\mathrm{fps}$ during steady-state operation, defined by stable film retraction height. We captured additional side-view recordings at 300 fps to observe rapid events such as film retraction, as shown in Figure \ref{fig: setup} (c). In this figure, we represent the puddle front contact angle $\theta_F$ and the rear contact angle $\theta_R$. Each experiment was repeated three times, with careful cleaning (isopropanol and deionised water) between trials.

\section{Flow Regimes on Smooth and Grooved Substrates}

\begin{figure}
    \centering
    \includegraphics[width=1\linewidth]{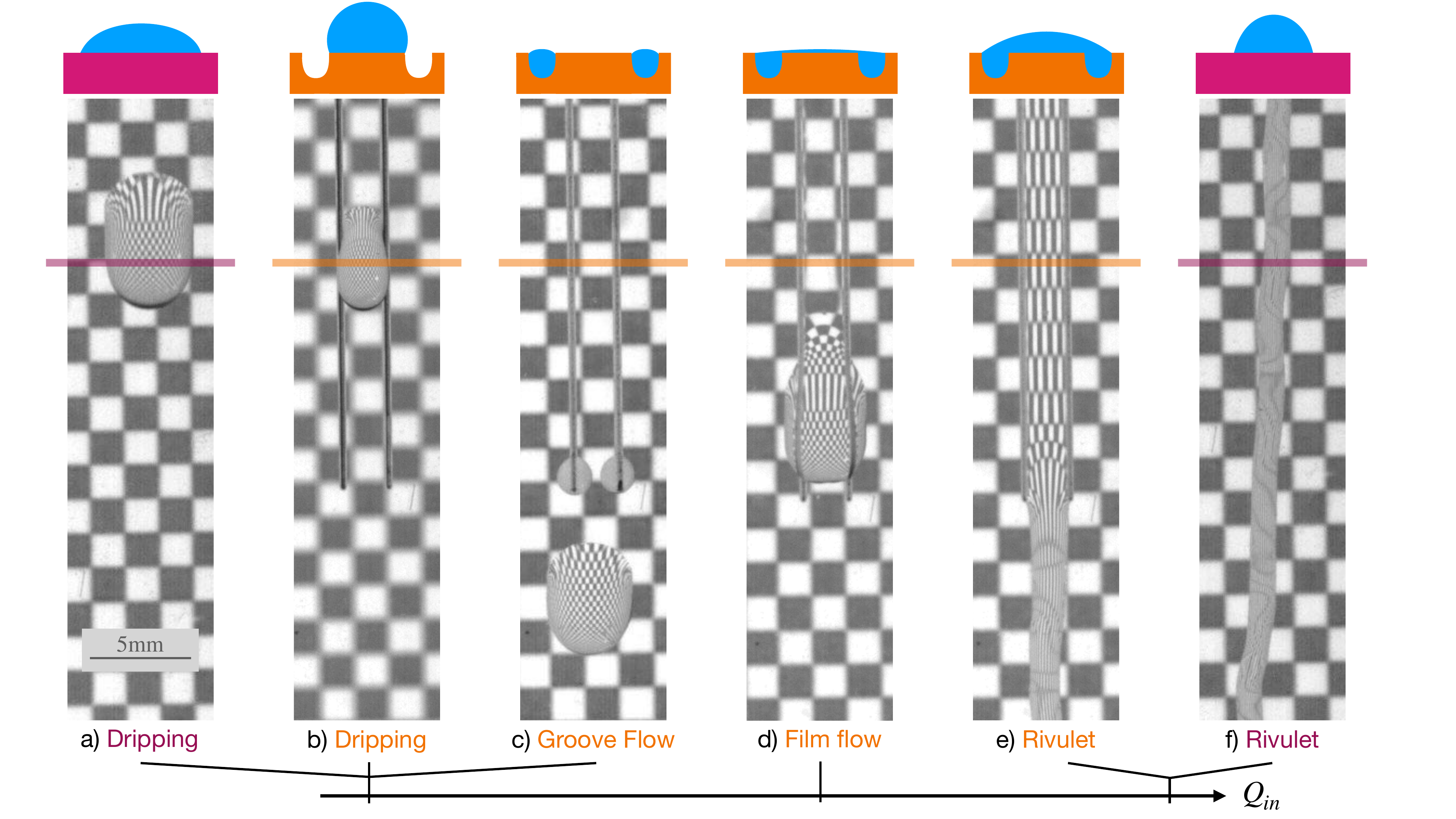}
    \caption{Sequence of flow regimes observed on vertical substrates as a function of flow rate $Q_{\mathrm{in}}$.
    Each image displays the fluid morphology for increasing $Q_{\mathrm{in}}$, with a checkered background that reveals film thickness via refraction: compressed patterns indicate higher curvature and thicker liquid layers. Schematics above each image illustrate the fluid cross-section at the location marked by the colored line in the corresponding photo. Smooth surfaces are shaded purple, and grooved cross-sections are shaded orange. At low flow rate on a smooth substrate, droplets form and slide once their weight overcomes contact angle hysteresis (a). At high flow rates, these droplets merge into a rivulet that eventually meanders due to inertial effects (f). On grooved substrates, new regimes emerge. When grooves are dry, they confine droplet width but do not transport liquid (b). Once prewetted, the same $Q_{\mathrm{in}}$ yields Groove Flow, where a puddle at the syringe tip feeds the grooves and droplets detach from the bottom (c). At a higher flow rate, a stabilised film stretches between the grooves and periodically ruptures at their termination, initiating the Film Flow regime (d). At very high flow, this film becomes a rivulet, anchored by the grooves and held in a straight path (e).}
    \label{fig: Q_behaviour}
\end{figure}

Because this system introduces entirely new dynamics, we first describe how the fluid interacts with the substrate, which is initially smooth and then grooved, for different flow rates $Q_{\rm{in}}$. This overview, summarised in Figure~\ref{fig: Q_behaviour}, highlights the richness of behaviours that emerge. Thin fluid layers are difficult to visualise directly, so we placed a checkered pattern between the light source and the substrate. Pattern deformation by refraction provides a qualitative view of the film: compressed pattern regions indicate greater interface curvature and correspond to a thicker fluid layer.

\subsection{Smooth}

Injecting water onto a smooth vertical surface results in two stable morphologies: droplets at low flow rates and rivulets at high ones, illustrated in purple in Figure~\ref{fig: Q_behaviour} (a) and (f). At low flow rates, the fluid accumulates into a droplet that remains pinned between the surface and the syringe. This pinning arises from contact angle hysteresis, the imbalance between the advancing and receding contact angles. As the droplet grows, fed by the continuous input, its weight eventually overcomes this capillary barrier, triggering motion~\cite{furmidge_sliding_1962,extrand_retention_1990,quere_drops_1998}. The droplet shape reflects its velocity: slow-moving droplets maintain an oval profile, while faster ones develop a rear corner or even a sharp cusp. At high speeds, the droplet rear becomes unstable and begins to shed smaller satellites, a regime known as pearling~\cite{podgorski_corners_2001,le_grand-piteira_shape_2005}.

As the flow rate increases further, this sequence of individual droplets gives way to a continuous stream, a rivulet. This transition is energetically favourable as a connected stream reduces the total interfacial energy~\cite{ghezzehei_constraints_2004}. Initially, the rivulet flows straight along the direction of gravity, stabilised by surface tension and contact line pinning. At even higher flow rates, however, inertial effects and local disturbances can destabilise the thread, leading to meandering. These two morphologies, droplets and rivulets, define the full range of stable behaviours accessible on a smooth vertical substrate. The transition between these two morphologies is at a flow rate about $Q_{\rm{in}}\approx 80 \,\rm{mm^3/s}$. A schematic diagram shows a cross-section of the system at the coloured line in the figure \ref{fig: Q_behaviour}.

\subsection{Two Vertical Grooves}

We now consider substrates patterned with a pair of vertical grooves, represented by orange schematics in Figure \ref{fig: Q_behaviour}. These grooves introduce new pathways for fluid motion, altering the formation, growth, and detachment of droplets. To get a feeling of the order of magnitude, we will consider the sample whose groove aspect ratio is $d/w=1.80$, corresponding to the most significant groove cross-section.

First, we apply a low flow rate of $Q_{\rm{in}} = 1.66 \,\mathrm{mm^3/s}$. When the grooves are dry, they act as barriers that constrain the droplet geometry. The base of the resulting droplet has a width equal to the spacing $s$ between the grooves, as illustrated on the schematic of Figure \ref{fig: Q_behaviour} (b). As the droplet exits the grooved zone, its shape relaxes into a rounder profile, and it slows down. This deceleration is consistent with what is observed for droplets on vertical strips, an analogous system, since dry grooves behave like geometric boundaries. In such systems, droplet velocity is inversely proportional to strip width \cite{bintein_kirigami_2023}. Therefore, even when dry, the grooves already influence droplet dynamics.

Once the grooves are prewetted by sliding droplets along them, injecting the same flow rate as before yields a markedly different behaviour. Instead of periodic detachment from the syringe tip, a small and stable puddle forms at the syringe–groove junction. This puddle maintains a constant volume and acts as a conduit that feeds water into both grooves, as represented in the schematic of Figure \ref{fig: Q_behaviour} (c). At the lower end of the substrate, two small droplets emerge, one per groove, and gradually grow until they merge and detach as a single droplet, visible in Figure~\ref{fig: Q_behaviour}(c). Immediately after detachment, two new droplets begin to form at the groove tips, initiating the cycle again. This dripping is decoupled from the syringe tip properties or position and relies only on groove characteristics. The grooves thus transition from passive boundaries (when dry) to active fluid conveyors, with the syringe-side puddle acting as a steady junction. For this reason, we refer to this regime as Groove Flow.

As the flow rate increases further, with $Q_{\rm{in}}\in[1.66,3.33]\,\mathrm{mm^3/s}$, the puddle at the syringe tip, previously stationary, begins to swell. Once it grows large enough, it detaches and slides down the grooves, colliding with and entraining the droplets already present at the lower ends. Importantly, even during this swelling and sliding, the grooves continue to transport water. The imposed flow rate $Q_{\rm{in}}$ is thus partitioned into two: one through the grooves and another into the syringe-side puddle. The threshold at which the puddle begins to swell depends on groove geometry. Intuitively, the larger the groove cross-section, the more liquid it can transport before reaching saturation.

When the input flow rate approaches $Q_{\rm{in}}\approx3\,\mathrm{mm^3/s}$, the dynamics shift once again. Occasionally, the detachment of the syringe-side puddle leaves behind a thin liquid sheet confined between the grooves. At first, this film is unstable and ruptures quickly, but with increasing flow rate, it stretches further before breaking. Eventually, we reach a critical threshold where the film stabilises along the entire groove length. This continuous sheet connects the syringe to the droplet below, laterally anchored by the grooves, represented in the schematic of Figure \ref{fig: Q_behaviour} (d).

\begin{figure}
    \centering
    \includegraphics[width=0.7\linewidth]{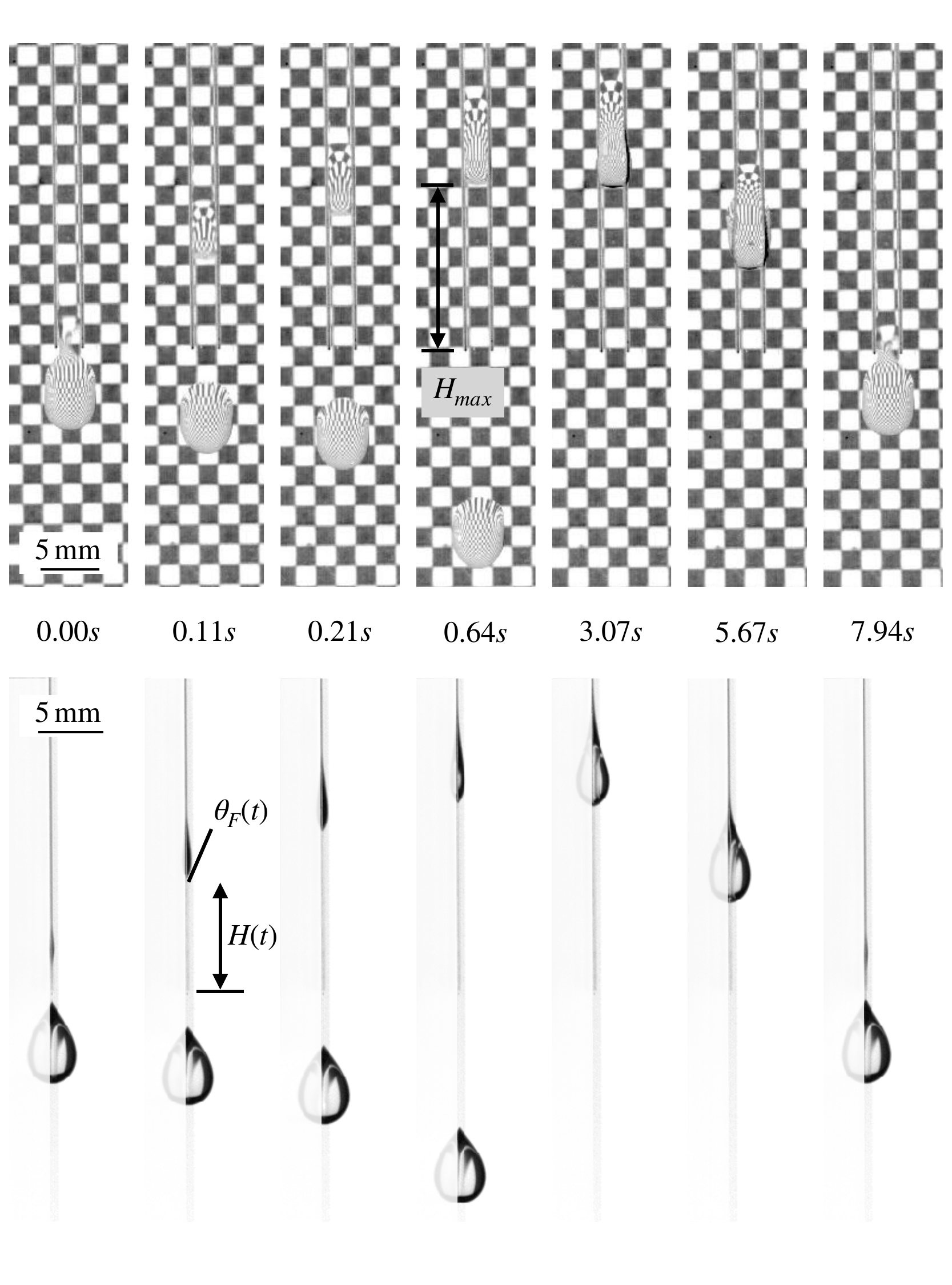}
    \caption{
Cyclic dynamics of film rupture and droplet release in the stabilised film regime.
Top and bottom rows show synchronised front and side views of the same experiment: groove spacing of $s = 2.25\,\mathrm{mm}$, an aspect ratio $d/w = 1.80$, and an input flow rate of $Q_{\mathrm{in}} = 3.33\,\mathrm{mm^3/s}$. The time elapsed between each pair of images is indicated.
(Top) A checkered pattern is placed between the backlight and the substrate to visualise film curvature via refraction: strong pattern deformation signals thicker fluid layers. Yet the film between the droplet and syringe is nearly imperceptible; the pattern remains barely distorted, highlighting how thin the film is. As the droplet slides and crosses the groove termini, at $t=0\,\rm{s}$, lateral confinement disappears and the film ruptures. The film retracts upward, forming at its base a new puddle that grows as the film pulls back. Retraction and early puddle growth occur within $0.64\,\mathrm{s}$. The puddle then pins at a stable retraction height $H_{\mathrm{max}}$, where capillary forces balance its weight, and continues to inflate for another $2.5\,\mathrm{s}$ (until $t = 3.07\,\mathrm{s}$). Once the puddle becomes heavy enough, it depins and slides downward over approximately $5\,\mathrm{s}$ (reaching $t = 7.94\,\mathrm{s}$). When the rear of the droplet reaches the end of the groove, the film ruptures again, restarting the cycle.
(Bottom) Side view of the same dynamics, emphasising puddle inflation and the evolution of the front contact angle $\theta_{\mathrm{F}}$, which increases steadily until the droplet detaches.} 
    \label{fig: Front-Side_View}
\end{figure}

This stabilised film marks the onset of the third regime, the Film Flow regime, observed for $Q_{\rm{in}}\in[3.33,80]\,\mathrm{mm^3/s}$ (Figure~\ref{fig: Q_behaviour} (d)). In this state, the droplet detaches and slides down between the grooves, leaving behind a continuous film that remains connected to the syringe. This anchored film maintains the water supply and prevents puddle swelling at the top. Once the droplet exits the grooved zone, the film loses its lateral confinement and breaks. Remarkably, this breakup does not occur chaotically: the sheet retracts upward. Because water continues to flow in, the film does not fully collapse up to the syringe tip. Instead, it retracts until a new puddle forms at a height $H_{\mathrm{max}}$, where gravitational weight balances surface tension. The system then enters a repeatable cycle, illustrated in Figure~\ref{fig: Front-Side_View} for $s=2.25\,\rm{mm}$, $d/w=1.8$ and $Q_{\rm{in}}=3.33\,\rm{mm^3/s}$. We first see a rapid film retraction accompanied by a growing puddle in half a second. Then the puddle stops at $H_{\mathrm{max}}$ and further grows during $2.5\,\rm{s}$. At that point, its weight overcomes capillary force and it begins to slide down for $5 \,\rm{s}$. Ultimately, the rear of the droplet reaches the end of the groove, leading to film breakup. What is striking in these figures is how imperceptible the film appears behind the puddle: the checkered pattern is barely disturbed, giving the impression that no film is present at all. Yet the smooth connection between the puddle and the film, together with the full sequence of events, leaves no doubt. The film is indeed there, stretched at a thickness that borders on the undetectable to the naked eye.

Within this regime, we identify two behaviours. At lower flow rates ($Q_{\rm{in}}\in[3.33,10]\,\mathrm{mm^3/s}$), the film repeatedly retracts and reforms, producing a well-defined cyclic instability governed by $H_{\mathrm{max}}$. At higher flow rates ($Q_{\rm{in}}\in[10,80]\,\mathrm{mm^3/s}$), the retraction height progressively diminishes until it vanishes. In this limit, droplets form directly from the continuous film at the groove tips, resembling remote dripping.

Finally, at very high flow rates, $Q_{\rm{in}}>80\,\mathrm{mm^3/s}$, droplet emission becomes so rapid that individual drops no longer drip. Instead, they merge into a continuous rivulet. Comparing the pattern density between grooves in Figures~\ref{fig: Q_behaviour} (d) and (e) reveals that increasing the flow rate also thickens the film. The transition toward rivulet for a grooved substrate occurs at a comparable flow rate to that on a smooth one, as shown in Figure~\ref{fig: Q_behaviour}(e) and (f). Yet the presence of grooves introduces a crucial difference. Within the grooved region, the rivulet remains perfectly linear, strictly aligned with the groove direction. The grooves act as lateral anchors, suppressing transverse capillary instabilities and enforcing a straight flow path, clearly visible in Figure~\ref{fig: Q_behaviour} (e). Once the rivulet exits the structured zone, however, its cross-section narrows under the pull of surface tension, further demonstrating how the grooves sustain a broader and thinner film.

The present study focuses exclusively on this Film Flow regime. Specifically, we examine the flow range $Q_{\mathrm{in}}\in\left[0.6,6.7\right]$, where the film remains stable between the grooves and where an apparent film retraction characterises the instability. Our goal is to study the morphology, stability, and retraction dynamics of the film in this regime. 

\section{Film thickness}
We first examine the stabilised portion of the film and look at how the imposed flow rate $Q_{\mathrm{in}}$ impacts its thickness $e$. Understanding this link is crucial for two reasons. It enables us to predict and control film thickness directly from substrate geometry and flow conditions. Second, this thickness plays a role in the cyclic breakup and reformation of the film.

\subsection{Results}

\begin{figure}
    \centering
    \includegraphics[width=1\linewidth]{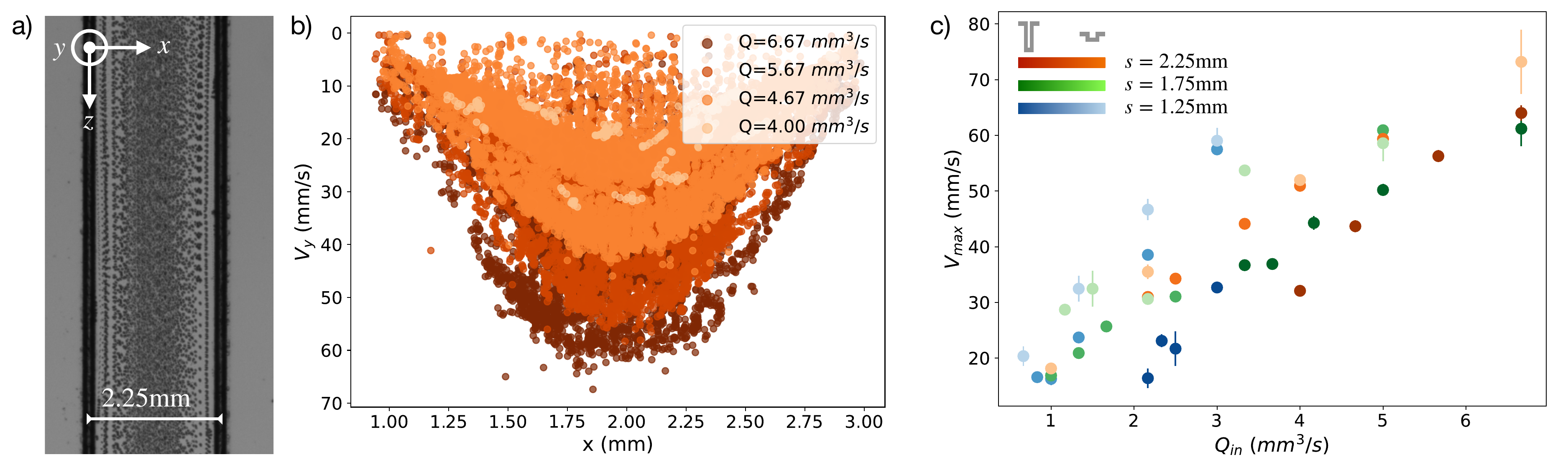}
    \caption{Film velocity measurements using particle tracking velocimetry (PTV).
    (a) Superposition of 2000 consecutive frames showing $20\,\mu\mathrm{m}$ tracer particles flowing through a thin film confined between two grooves spaced by $s=2.25\,\mathrm{mm}$ with $Q_{\rm{in}} = 6.67\,\rm{mm^3/s}$. Tracers cluster near the centre, indicating faster flow in this region.
    (b) Reconstructed horizontal velocity profiles across the film, exhibiting a parabolic shape consistent with confined Stokes flow. The apparent “filled” profiles reflect the 3D nature of the flow: particles occupy different heights, resulting in multiple velocities being detected at each horizontal position.
    (c) Maximum velocity $V_{\mathrm{max}}$, averaged from the 10 fastest particles, plotted as a function of $Q_{\mathrm{in}}$ for various groove spacings $s$ and geometries $d/w$. Velocity increases with $Q_{\mathrm{in}}$ and reveals two trends: (i) at fixed groove geometry (same shade), smaller spacings (blue) yield higher $V_{\mathrm{max}}$ due to thicker films; (ii) at fixed spacing (same color), deeper grooves (darker tones) reduce $V_{\mathrm{max}}$, consistent with flow being diverted into the grooves.}
    \label{fig: PTV}
\end{figure}

Our goal is to quantify film thickness; however, direct optical techniques are poorly suited for our system. Interferometry loses resolution in the $50$–$100\,\mu\mathrm{m}$ range, while profilometry requires reflectivity or scattering contrast, which transparent water films on acrylic lack. Instead, we use particle tracking velocimetry (PTV), a robust and non-invasive method. By measuring internal velocity fields and applying a Stokes-flow model, we infer thickness indirectly yet reliably.

Figure~\ref{fig: PTV} summarises the velocity field measured within the stabilised film using PTV (see Experimental Setup). In a typical flow illustrated in Figure~\ref{fig: PTV} (a), tracers concentrate near the film centre, indicating faster motion in this region. The reconstructed velocity profiles, plotted in Figure \ref{fig: PTV}, show a parabolic shape, with maximum speed at the centre and slower flow near the edges, reflecting the confinement imposed by the groove boundaries. The apparent “filled” character of the profiles arises from the three-dimensional nature of the flow: tracer particles are distributed across different heights within the film, so multiple velocities are detected at a given horizontal position. As the imposed flow rate $Q_{\mathrm{in}}$ increases, the maximum velocity within the film increases accordingly, consistent with the expected scaling between flow rate and film thickness.

We then average the speed of the 10 fastest particles for various groove geometries and spacings and plotted them as a function of the imposed flow rate $Q_{\mathrm{in}}$ in Figure \ref{fig: PTV} (c). As expected, $V_{\mathrm{max}}$ increases with $Q_{\mathrm{in}}$. Two additional trends emerge. First, at a given flow rate and fixed groove geometry (same shade), smaller spacings (blue points) yield higher velocities than larger ones (orange points). When the film is stretched across a wider spacing, its thickness decreases, which in turn reduces the maximum speed that can be sustained by the film. Second, at fixed spacing (same colour), we observe that deeper grooves (darker tones) result in lower maximum velocities. This suggests that more fluid is diverted into the grooves, leaving less available to form the film, resulting in a thinner film and reduced flow speed.

\subsection{Model}

We now develop two complementary approaches to estimate the film thickness. The first method relies on velocity profiles extracted from PTV measurements, while the second predicts thickness based on the imposed flow rate and substrate geometry. Comparing both models allows us to validate the theoretical scaling and refine its predictive accuracy.

In the first approach, we assume that inertial effects are negligible ($\mathrm{Re} < 10$), so the flow within the film can be described as a gravity-driven Stokes flow. The motion of thin liquid films is primarily governed by viscous friction, which renders the lubrication approximation particularly suitable~\cite{gennes_gouttes_2013}. The balance between gravity and viscosity yields
\begin{equation}
\rho g = \eta \dv[2]{V}{y},
\end{equation}
where $V(y)$ is the out of plane velocity, $\rho$ the fluid density, $\eta$ the dynamic viscosity, and $y$ the height perpendicular to the substrate. Solving this equation with a no-slip condition at the substrate and a zero-shear condition at the air–liquid interface gives the velocity profile
\begin{equation}
V(y) = \frac{\rho g}{\eta} \left(ey - \frac{y^2}{2} \right),
\label{eq:film_speed}
\end{equation}
where $e$ denotes the film thickness. The maximum velocity occurs at the free surface, $y = e$, leading to:
\begin{equation}
V_{\mathrm{max}} = \frac{\rho g e^2}{2\eta}.
\label{eq:stockes}
\end{equation}
This relation provides a way to estimate the film thickness from the measured peak velocity
\begin{equation}
e_{\mathrm{PTV}} = \sqrt{ \frac{2\eta V_{\mathrm{max}}}{\rho g} }.
\label{eq:ePTV}
\end{equation}

To build a predictive model based solely on the imposed flow rate, we consider the total input as a combination of flow through the film and through the grooves
\begin{equation}
Q_{\mathrm{in}} = Q_f + 2Q_g,
\label{eq: saturation}
\end{equation}
where $Q_f$ is the flow rate through the film and $2Q_g$ represents the maximum flow that can be entirely transported through the grooves. This saturation threshold $2Q_g$ is determined experimentally in the Groove Flow regime by gradually increasing $Q_{\mathrm{in}}$ until a visible puddle begins to grow at the syringe–groove junction. We find that $2Q_g$ depends on groove geometry: for shallow grooves ($d/w = 0.54$), $2Q_g = 0.08 \pm 0.02\,\mathrm{mm^3/s}$; for intermediate grooves ($d/w = 1.13$), $2Q_g = 0.42 \pm 0.05\,\mathrm{mm^3/s}$; and for deep grooves ($d/w = 1.80$), $2Q_g = 1.67 \pm 0.17\,\mathrm{mm^3/s}$.

We then express the film flow rate $Q_f$ as the product of cross-sectional area and average velocity: $Q_f = es\widetilde{V}$, with $s$ the groove spacing and $\widetilde{V}$ the mean velocity in the film, assuming uniform thickness. Integrating Equation~\eqref{eq:film_speed} over the film thickness gives $\widetilde{V} = \rho g e^2 / 3\eta$. Substituting this into the expression for $Q_f$ yields
\begin{equation}
e_{\mathrm{flow}} = \left( \frac{3\eta (Q_{\mathrm{in}} - 2Q_g)}{\rho g s} \right)^{1/3}.
\label{eq:eflow}
\end{equation}

A comparison between the experimental method $e_{\rm{PTV}}$ and theoretical one $e_{\rm{flow}}$ is presented in Figure~\ref{fig: e(Q)} (a). While the predicted and measured values are strongly correlated, a systematic offset is observed. This is corrected by introducing an approximated prefactor $\xi = 4/5$, such that
\begin{equation}
e_{\mathrm{flow}} = \xi\, e_{\mathrm{PTV}}.
\end{equation}
This correction accounts for model simplifications, such as the assumption of infinite film width. In reality, the film is confined between two grooves and exhibits curvature at the edges, which reduces the effective flow area. Incorporating this correction, the final expression for film thickness becomes
\begin{equation}
e_{PTV} = \frac{1}{\xi} \left( \frac{3\eta}{\rho g} \frac{Q_f}{s} \right)^{1/3}.
\label{eq: thickness}
\end{equation}
This final expression provides a compact relationship between film thickness, flow parameters and groove geometry. The measured maximum film thickness was consistently greater than $50\,\mu\mathrm{m}$, whereas thin-film instability theory predicts a critical thickness on the order of $1 \rm{\mu m}$ \cite{mezic_stability_1998}. This discrepancy can be understood by noting that our model assumes a uniform film thickness, while in reality the confined geometry imposes a parabolic profile. Consequently, the thickness near the edges can approach values comparable to the limit of the thin-film instability.

\subsection{Discussion}

\begin{figure}
    \centering
    \includegraphics[width=1\linewidth]{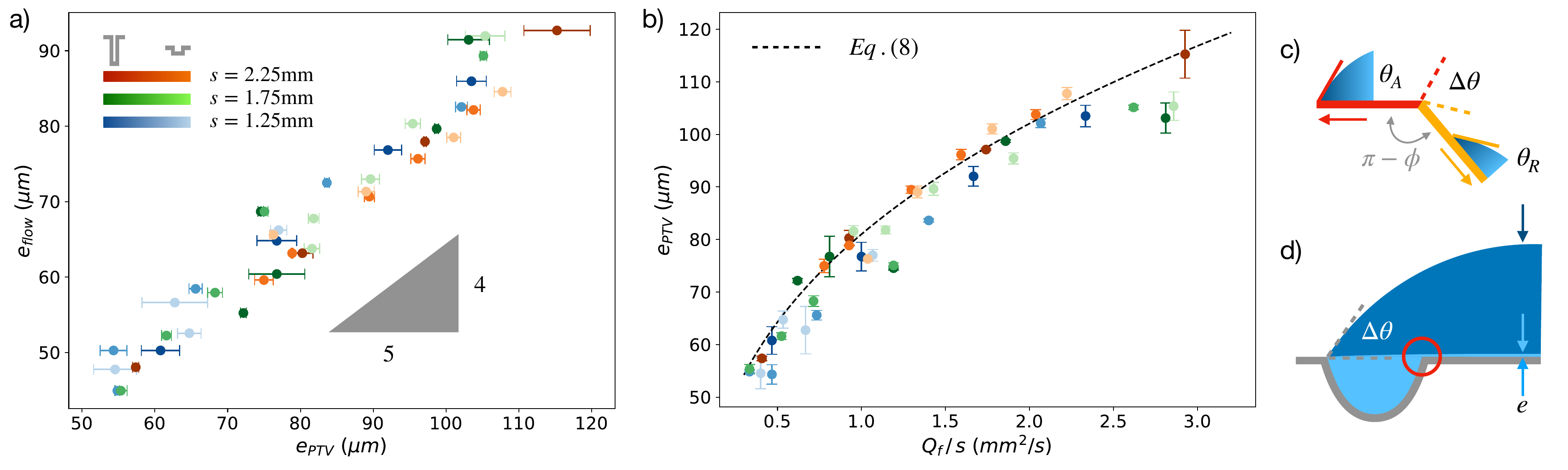}
    \caption{Film thickness measurements and model validation.
    (a) Comparison between experimental film thickness $e_{\mathrm{PTV}}$, inferred from particle tracking velocimetry (Equation \ref{eq:ePTV}), and the theoretical prediction $e_{\mathrm{flow}}$ based on a uniform Stokes flow model (Equation \ref{eq:eflow}). A systematic offset (slope $\xi = 4/5$) is attributed to the finite width of the film, which reduces the effective flow area.
    (b) Collapse of all measured film thicknesses $e_{\mathrm{PTV}}$ as a function of scaled film flow rate $Q_f/s$. This collapse confirms the model central assumption: once groove flow saturates, the remaining input is redirected entirely into the intergroove film (Equation~\ref{eq: saturation}). The dashed line represents the predicted model (Equation~\ref{eq: thickness}), which is validated across various groove geometries and spacings.
    (c) Schematic showing how a corner enables the triple line to accommodate a range of contact angles $\Delta\theta$, enhancing pinning and film stability.
    (d) Vertical cross-section of the outer groove edge. Despite thinning at low $Q_{\mathrm{in}}$, the film remains pinned as the contact angle stays within the hysteresis range (grey dashed lines). The red circle highlights a region of steep velocity gradients between film and groove flow, where small perturbations may trigger dewetting.}
    \label{fig: e(Q)}
\end{figure}

Figure~\ref{fig: e(Q)}(b) compares the experimental film thickness $e_{\mathrm{PTV}}$ (Equation~\ref{eq:ePTV}) with the scaled film flow rate $Q_{\mathrm{f}}/s$. Remarkably, data from all groove geometries and spacings collapse onto a single curve. This collapse supports the central assumption of the model: once groove flow saturates, the remaining input is entirely redirected into a stabilised intergroove film (Equation~\ref{eq: saturation}). The dashed line shows the theoretical prediction $e_{\mathrm{flow}}$ based on a confined Stokes flow model (Equation~\ref{eq: thickness}). The strong agreement across conditions confirms the validity of the scaling and demonstrates that film thickness can be reliably inferred from the input flow rate and groove geometry alone. The resulting films can be extremely thin, down to $50\,\mu\mathrm{m}$, despite being composed of pure water, without surfactants. Such thicknesses are typically unstable on smooth surfaces, yet here, they are robust and fully tunable via syringe flow rate alone. 

Still, we note that no direct thickness measurements were made for reasons explained earlier. Our model assumes an ideal Stokes flow with uniform thickness. In reality, the flow cross-section is not rectangular but curved near the groove edges. This lateral curvature reduces the effective flow area, resulting in a slight overestimation of the average velocity in the model. As a result, the actual film thickness is likely even thinner than predicted, an outcome that further underscores the ability of grooves to support thin water films. 

The next question is why the liquid film remains so stable. On a completely smooth surface, water films of similar thickness are prone to rupture or instabilities. Here, however, the film resists destabilisation over surprisingly long distances. To understand this behaviour, it is helpful to consider the role of edge pinning, illustrated by the corner shown in Figure~\ref{fig: e(Q)}(c).

Consider a droplet moving leftward along a red surface. To advance, its front contact angle must reach $\theta_{\mathrm{F}}$. Now imagine a droplet moving rightward on a yellow surface; its rear contact angle will take the value $\theta_{\mathrm{R}}$. Finally, suppose the droplet is located on the yellow surface with its contact line precisely at the corner where the yellow and red surfaces meet with an angle $\phi$. For the droplet to advance onto the red surface, its front angle must match $\theta_{\mathrm{F}}$, and to recede on the yellow surface, its rear angle must match $\theta_{\mathrm{R}}$. Thus, when the triple line sits at the corner, it can remain pinned across a wide range of angles $\Delta\theta$, accommodating changes in droplet volume without moving. The total angular range allowed by the corner is given by $\Delta\theta = (\theta_{\mathrm{A}} - \theta_{\mathrm{R}}) + \phi$, where $\phi$ is the corner angle. A sharper corner extends this range, making pinning even more robust.

This principle explains the behaviour observed at the outer edge of the grooves. As the flow rate increases, the film thickens according to Equation~\eqref{eq: thickness}, and the contact angle at the film edge rises. As long as this angle remains within the contact angle hysteresis range, the triple line stays pinned at the groove corner, and the film remains laterally confined without overflowing. Conversely, for an extremely thin film, the receding contact angle is very small, below the horizontal. However, since the grooves remain filled in the Film Flow regime, the contact angle at the triple line never reaches the receding threshold $\theta_{\mathrm{R}}$, and depinning does not occur. As a result, the groove edge acts as a robust pinning site, enabling the film to remain stable over a wide range of thicknesses.

But what happens when the film finally destabilises? In practice, this typically occurs at the film extremities, whether the syringe side or the puddle one. The precise mechanism remains difficult to capture, but a likely hypothesis involves the inner edges of the grooves as illustrated in Figure \ref{fig: e(Q)} (d) with a red circle. These locations concentrate velocity gradients between the film and the groove. A small perturbation, especially under low-flow conditions, may be sufficient to trigger dewetting from these points, initiating whole-film rupture.

\section{Periodic Film Retraction}

\begin{figure}
    \centering
    \includegraphics[width=1\linewidth]{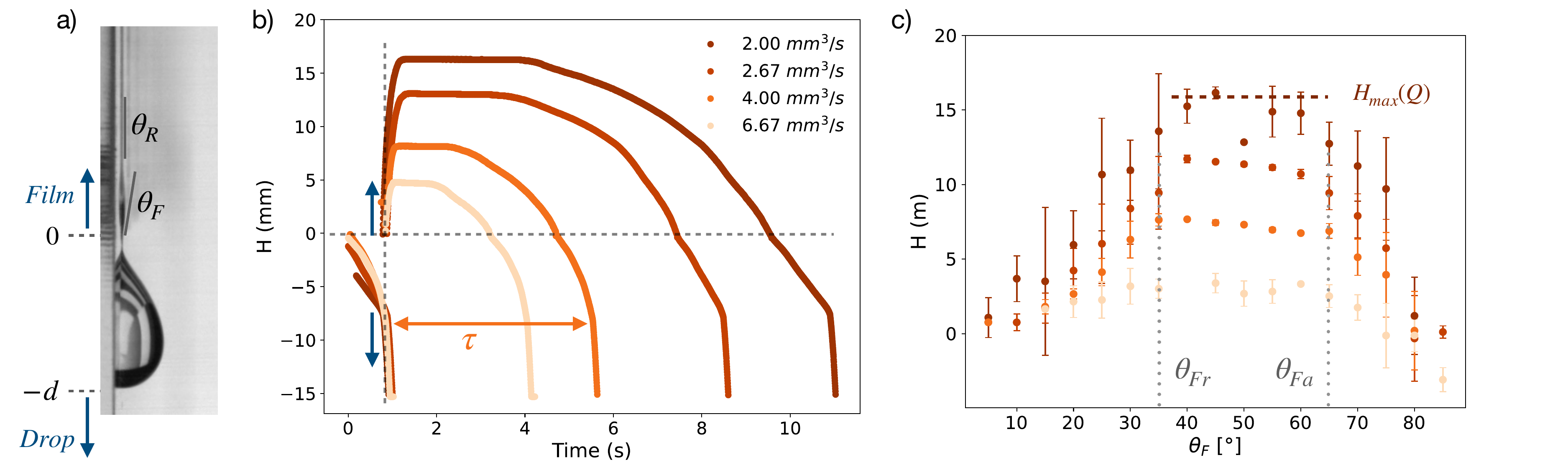}
    \caption{Cyclic dynamics of film retraction, puddle inflation, and droplet release.
        (a) Representative frame showing the start of a new cycle, triggered when the rear of the droplet passes the groove termination ($H = 0$). At this point, the film loses its lateral confinement, ruptures, and retracts upward. A new puddle forms at the rupture point and inflates as the film retracts. It is characterised by a front contact angle $\theta_{\mathrm{F}}$ and a vanishing rear angle $\theta_{\mathrm{R}} = 0^\circ$, due to the smooth connection with the film.
        (b)  Time evolution of the puddle lower boundary $H(t)$ as a function of time, for various flow rates on the substrate $d/w=1.80$ and $s=2.25\,\rm{mm}$. All trajectories follow the same sequence: rapid retraction, a plateau during which the puddle inflates, sliding, and detachment. Kinks at $H = 0$ and $H = -d$ correspond to the passage of the front and rear across the groove termination. Lower $Q_{\mathrm{in}}$ leads to taller retractions and longer inflation times. The cycle period $\tau$ is defined as the time between two successive film ruptures; the following cycle is not shown.
        (c) Evolution of the front contact angle $\theta_{\mathrm{F}}$ over one cycle. After rupture, $\theta_{\mathrm{F}}$ increases with puddle growth until reaching the receding threshold $\theta_{\mathrm{Fr}} \approx 35^\circ$, where the contact line pins. As the angle continues to increase, it reaches the critical threshold of around $65^\circ$, at which point the droplet slides. Dashed lines indicate these thresholds. Only $\theta_{\mathrm{F}}$ is measured; the rear angle is assumed to be $\theta_{\mathrm{R}} = 0^\circ$ due to the smooth film connection.
        }
    \label{fig: droplet_retraction}
\end{figure}

While grooves provide robust lateral pinning throughout the confined region, this stabilisation is inherently local. At the downstream end, where the grooves terminate and lateral confinement abruptly vanishes, the film loses its geometric anchors. This discontinuity acts as a passive trigger for destabilisation. We now turn to this stage, where the system enters a self-sustained cycle of droplet inflation, detachment, and film retraction.

\subsection{Cycle description}

To characterise the cyclic dynamics in the stabilised film regime, we track the vertical motion of the droplet’s lower boundary across flow rates $Q_{\mathrm{in}}$ (Figure~\ref{fig: droplet_retraction}(a–c)). The cycle begins when the rear of the droplet crosses the groove termination ($H = 0$ in Figure~\ref{fig: droplet_retraction}(a)). At this moment, the front continues its descent, while the thin connecting film ruptures and retracts upward. A puddle forms at the breaking point, marked by a low front angle $\theta_{\mathrm{F}}$, and initiates the next cycle.

Figure~\ref{fig: droplet_retraction}(b) details this process in time. At $t=0$, the front has already exited the grooves, while the rear remains connected by a thin film. When the rear reaches the groove end, the film breaks. Two paths then emerge: one traces the front as it slides downward, the other follows the upward retraction of the film. This retraction occurs in two stages: a rapid phase driven by surface tension, then a plateau where the retraction front pins and the puddle inflates, constantly fed by $Q_{\rm{in}}$. Once the puddle is heavy enough, the lower contact line depins and sliding begins. The descent itself unfolds in stages. The droplet initially accelerates, then slows when the front reaches $H = 0$ (bottom of the grooves), producing a kink in the trajectory. It then moves at nearly constant velocity until the rear exits the grooves at $H=-d$, where a second kink marks renewed acceleration before complete detachment. Across flow rates, this sequence is preserved: lower $Q_{\mathrm{in}}$ produces taller, faster retractions and longer static inflation phases, while higher $Q_{\mathrm{in}}$ shortens the cycle. Once the droplet has fully exited, trajectories converge to a similar acceleration profile.

While these trajectories reveal when transitions occur, they do not explain what controls them. To answer that, we examine the front contact angle $\theta_{\mathrm{F}}$ throughout the cycle (Figure~\ref{fig: droplet_retraction}(c)). In the retraction phase, both front angle and height rise together until $\theta_{\mathrm{F}}$ reaches about $35^\circ$. At this point, the puddle pins: its height freezes while $\theta_{\mathrm{F}}$ keeps increasing as water accumulates. We define this threshold as $\theta_{\mathrm{Fr}}$, with $r$ for “receding.” The terminology echoes classical hysteresis tests, where a droplet recedes once its angle falls below a critical value. Here, the logic is inverted: below $\theta_{\mathrm{Fr}}$ the line recedes, but once it reaches $\theta_{\mathrm{Fr}}$ it abruptly pins. A second threshold appears near $65^\circ$, when the dynamics reverse. The contact angle continues to grow, but the puddle height decreases as the droplet begins to slide. We label this $\theta_{\mathrm{Fa}}$, with $a$ for “advancing,” mirroring classical inflation tests where motion starts at an advancing angle. Beyond this point, the droplet keeps gaining mass from the inflow and accelerates as it leaves the grooves, which pushes $\theta_{\mathrm{F}}$ toward $90^\circ$ at exit.

Despite large variations in $Q_{\mathrm{in}}$, these thresholds are remarkably consistent. Retraction always halts at $\theta_{\mathrm{Fr}}\approx35^\circ$, while sliding always starts at $\theta_{\mathrm{Fa}}\approx65^\circ$. Both are slightly lower than advancing/receding values measured on smooth horizontal substrates. This discrepancy may reflect the influence of the grooves, either during retraction by confining the puddle or during descent by guiding the droplet. Together, $\theta_{\mathrm{Fr}}$ and $\theta_{\mathrm{Fa}}$ provide robust geometric markers that govern the puddle cycle. 

\subsection{Results}

\begin{figure}
    \centering
    \includegraphics[width=1\linewidth]{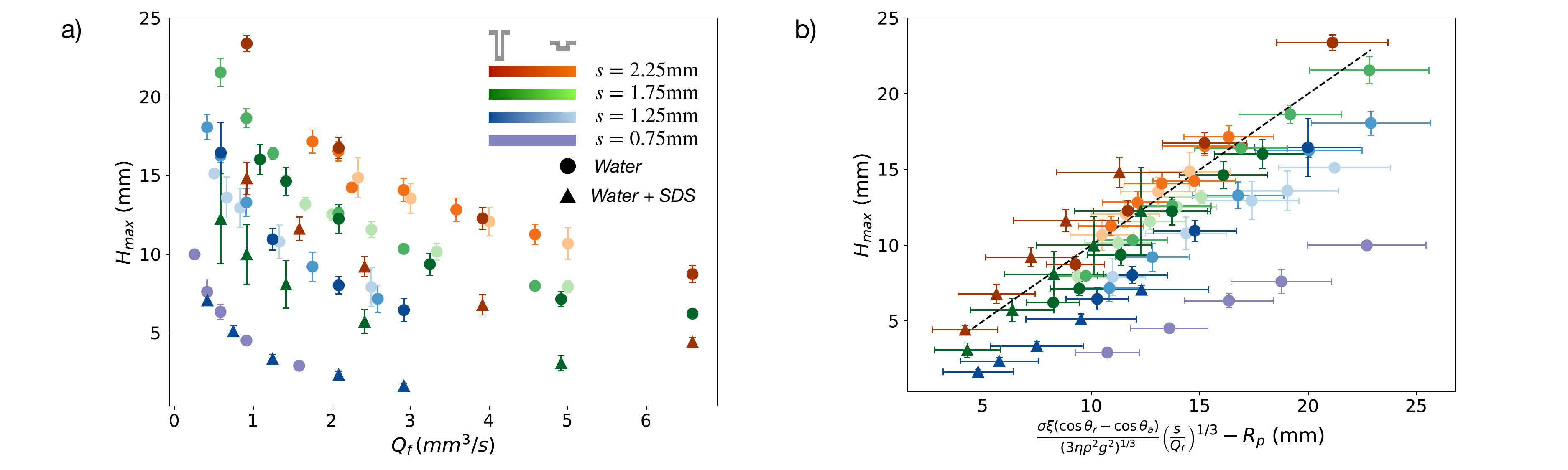}
    \caption{Retraction height as a function of film flow rate and model validation.
    (a) Experimental measurements of the maximum retraction height $H_{\mathrm{max}}$ as a function of film flow rate $Q_{\mathrm{f}} = Q_{\mathrm{in}} - 2Q_{\mathrm{g}}$. Each colour represents a different groove spacing $s$; darker shades correspond to deeper grooves (larger $d/w$). Filled circles indicate pure water, while filled triangles correspond to SDS-added water (lower surface tension). Two trends emerge: increasing $Q_{\mathrm{f}}$ reduces $H_{\mathrm{max}}$, as thicker films reach weight–capillarity balance more quickly; meanwhile, increasing $s$ increases $H_{\mathrm{max}}$, as wider films support stronger upward surface tension forces. SDS data (triangles) shift downward, consistent with a reduced surface tension lowering the capillary force required for balance. (b) Experimental data plotted as function of the model $H_{\mathrm{max}}$ based on a gravity–capillarity force balance (Equation~\ref{H_max}). The dashed line of slope 1 confirms the good agreement. However, at the smallest spacing $s = 0.75\,\mathrm{mm}$, the model overpredicts $H_{\mathrm{max}}$, likely because groove geometry begins to influence drainage and curvature, violating the flat-film assumption.}
    \label{fig: sheet_height}
\end{figure}

Figure~\ref{fig: sheet_height}(a) presents the experimentally measured retraction height $H_{\mathrm{max}}$ as a function of the film flow rate $Q_{\mathrm{f}} = Q_{\mathrm{in}} - 2Q_{\mathrm{g}}$. Each colour corresponds to a different groove spacing $s$, while the shade within each colour encodes groove aspect ratio: darker tones represent deeper grooves (larger $d/w$). For a given $s$, data points are ordered left to right by increasing $Q_{\mathrm{in}}$. Two key trends emerge. First, for a fixed spacing $s$, the retraction height $H_{\mathrm{max}}$ decreases with increasing flow rate. Second, for a fixed $Q_{\mathrm{f}}$, the retraction height increases with groove spacing. Finally, data for identical spacing $s$ but different groove aspect ratios superimpose on one another. This supports the relevance of $Q_{\mathrm{f}}$ as a unifying parameter: once the grooves saturate, $Q_{\rm{in}}>2Q_g$, any additional injected flow is routed through the film, and its effect on $H_{\mathrm{max}}$ depends only on the intergroove flow $Q_f$. Importantly, experiments performed with SDS (filled triangles) show a systematic downward shift compared to pure water (filled circles). Since SDS lowers surface tension, this shift suggests that $H_{\mathrm{max}}$ is sensitive to capillary effects. Altogether, these observations suggest that $H_{\mathrm{max}}$ depends on both the weight of the accumulated water and the surface tension pulling it upward. In the next section, we develop a theoretical model to capture this behaviour quantitatively.

\subsection{Modelisation}

At the end of the retraction phase, the puddle pins at its maximum height. We interpret this as the point where surface tension balances gravitational weight. The capillary force is approximated as
\begin{equation}
    F_{\sigma} = \sigma s \left( \cos\theta_R - \cos\theta_F \right),
\end{equation}
where $ \sigma $ is the surface tension, $ s $ the groove spacing, and $ \theta_{\mathrm{R}} $ (resp. $ \theta_{\mathrm{F}} $) the puddle rear (resp. front) contact angles. We use $ \theta_R = 0^\circ $ to reflect the smooth rear connection with the film, and set $ \theta_F = \theta_{\mathrm{Fr}} $, the front contact angle at the end of the retraction phase, measured experimentally. For water, we find $ \theta_{\mathrm{Fr}} = 38.5^\circ \pm 3^\circ $; for SDS, $ \theta_{\mathrm{Fr}} = 34^\circ \pm 6^\circ $. 
The opposing force arises from the mass of the puddle. We assume that it corresponds to mass of the film over a length $H_{\mathrm{max}} + R_p $, with $ R_p $ denoting half the puddle height. This leads to the expression
\begin{equation}
    F_g = \rho g e s (H_{\mathrm{max}} + R_p),
\end{equation}
where $ g $ the gravitational acceleration. The puddle height remains approximately constant across experiments, with $ 2R_p = 7.0 \pm 0.4\,\mathrm{mm} $.
Equating the two forces yields
\begin{equation}
    \sigma s \left( 1 - \cos\theta_{\mathrm{Fr}} \right) = \rho g e s \left( H_{\mathrm{max}} + R_p \right).
    \label{eq:force_balance}
\end{equation}
To express the film thickness $ e $ in terms of controllable parameters, we substitute Equation~\eqref{eq: thickness} (film thickness vs flow rate)
\begin{equation}
    e = \frac{1}{\xi}\left( \frac{3 \eta (Q_{\mathrm{in}} - 2Q_g)}{\rho g s} \right)^{1/3}.
\end{equation}
Inserting this into the force balance yields the final expression for the maximum retraction height:
\begin{equation}
    H_{\mathrm{max}} = \frac{\sigma \xi \left( 1 - \cos\theta_{\mathrm{Fr}} \right)}{(3 \eta \rho^2 g^2)^{1/3}} \left( \frac{s}{Q_{\mathrm{f}}} \right)^{1/3} - R_p.
    \label{H_max}
\end{equation}

This model predicts that retraction height increases with groove spacing and surface tension, and decreases with inter-groove flow rate.

\subsection{Discussion}

Figure~\ref{fig: sheet_height}(b) compares this theoretical prediction with the experimental data. The agreement is striking: across groove geometries, spacings, and fluid compositions, the data collapse onto a single straight line when plotted as a function of predictive model (Equation \ref{H_max}). This collapse confirms that the retraction height $H_{\mathrm{max}}$ is primarily governed by a balance between gravity and capillarity. The model captures the observed trends: $H_{\mathrm{max}}$ decreases with increasing flow rate (via thicker films) and increases with groove spacing (via stronger capillary support). It also predicts the downward shift caused by the addition of SDS, reflecting reduced surface tension. 

Despite its success, the model rests on several simplifying assumptions that begin to break down at the extremes. At the smallest groove spacing tested ($s = 0.75\,\mathrm{mm}$), the model systematically overpredicts $H_{\mathrm{max}}$. This suggests that when the film becomes tightly confined, the groove can no longer be treated as a passive boundary. As the groove cross-section becomes comparable to that of the film, it may actively participate in drainage by altering the way the film retracts. These effects alter the dynamic in ways not captured by our flat-film approximation. At the opposite extreme, our data approach but do not exceed the capillary length $\lambda = 2.67\,\mathrm{mm}$, beyond which Rayleigh–Plateau instabilities are expected to appear. In this regime, continuous wetting becomes difficult to maintain, as the $0.8\,\mathrm{mm}$ diameter nozzle struggles to bridge wide groove spacings reliably.

Another limitation lies in the assumption of a static force balance. In practice, the puddle continues to grow as it retracts, accumulating additional fluid from the still-active film. While this inflow is likely small compared to the final puddle mass, it introduces a temporal component that could subtly shift $H_{\mathrm{max}}$, especially if we consider higher input flow rates $Q_{\mathrm{in}}$.

In summary, the model captures the essential physics driving the retraction height, a gravity–capillarity balance shaped by groove spacing, film thickness, and surface tension. Its success demonstrates that $H_{\mathrm{max}}$ provides a robust, experimentally accessible observable for probing thin-film dynamics. Yet at the limits of groove spacing, confinement, or stability, additional mechanisms such as groove–film decoupling, time-dependent feeding, and curvature effects may emerge. These warrant further investigation in future work.

\section{Droplet Release}

\begin{figure}
    \centering
    \includegraphics[width=1\linewidth]{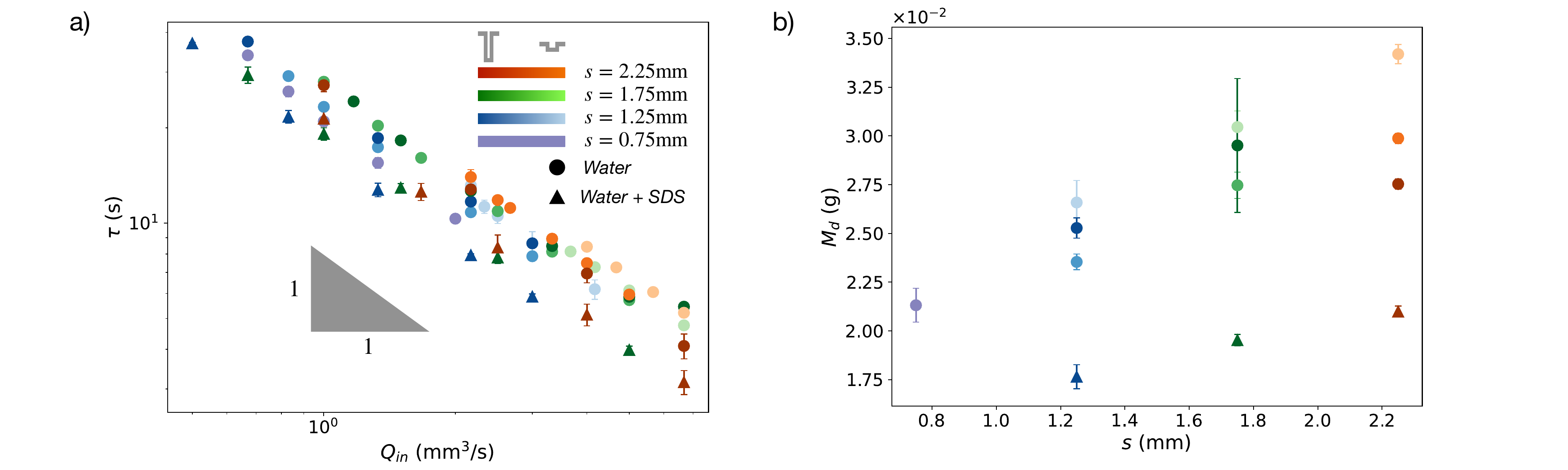}
    \caption{(a) Log-log plot of the average cycle duration $\tau$ as a function of imposed flow rate $Q_{\mathrm{in}}$, compiled across groove geometries ($d/w$), spacings ($s$), and fluid compositions (pure water vs. SDS). Surfactant addition reduces $\tau$ across all conditions, while increasing groove spacing tends to delay detachment.
    (b) Corresponding droplet mass $M_d = \rho Q_{\mathrm{in}} \tau$ as a function of groove spacing $s$. Wider groove spacing tends to produce larger droplets, while SDS reduces droplet mass across all spacings.
}
    \label{fig: tau_m}
\end{figure}

We now turn to the detachment of the droplet following each film rupture. We measure the time between successive film ruptures $ \tau$, and from this, infer the mass of the released droplet. The duration of one cycle $\tau$ is represented in Figure \ref{fig: droplet_retraction} (b) for one of the flow rates.

Figure~\ref{fig: tau_m}(a) presents the average cycle duration $\tau$ as a function of the imposed flow rate $Q_{\mathrm{in}}$, compiled across all experiments varying groove geometry ($d/w$), spacing ($s$), and fluid composition (pure water vs. SDS). All data align along parallel lines of slope $-1$ in a log–log plot, revealing a robust inverse relation
\begin{equation}
\tau \propto Q_{\mathrm{in}}^{-1}.
\end{equation}
This relationship confirms that the rate of mass accumulation governs the timescale of droplet formation: at lower flow rates, it takes longer to build a droplet large enough to detach. Here, unlike previous regimes where only the film flow rate $Q_f$ mattered, the total input flow $Q_{\mathrm{in}}$ sets the pace, because the entire injected volume is redirected toward the droplet, which becomes the system sole sink.

Despite this scaling, secondary trends emerge. For pure water, increasing the groove spacing $s$ leads to longer cycle durations, suggesting that wider films provide stronger capillary resistance and require more time to reach detachment. Conversely, SDS reduces $\tau$ across all spacings and geometries. The surfactant thus accelerates the release cycle by lowering the surface tension. As a consequence, $\tau$ is a function of both spacing $s$ and surface tension $\sigma$. Assuming that all incoming flow contributes to droplet growth between ruptures, the droplet mass can be directly estimated as
\begin{equation}
M_d = \rho Q_{\mathrm{in}} \tau(s,\sigma).
\end{equation}

Figure~\ref{fig: tau_m}(b) presents the measured droplet mass $M_d$ as a function of groove spacing $s$. Larger spacings consistently yield heavier droplets, while the addition of SDS reduces retention across all spacings. These observations suggest that droplet release arises from a dynamic balance between gravity and capillarity. Unlike the static force balance governing retraction height, here the competition unfolds over time. Wider groove spacings increase the film width, strengthening surface tension forces and delaying release. Lowering surface tension with SDS promotes earlier detachment. Both effects modulate the critical mass needed to initiate sliding and, therefore, set the final droplet mass.

In summary, this final stage completes the picture of the destabilisation cycle. While $Q_{\mathrm{in}}$ controls the rhythm of droplet formation, groove spacing and surface tension fine-tune the release.

\section{Summary and Outlook}

We demonstrated that geometry alone, specifically, a pair of laser-engraved grooves, is sufficient to stabilise, shape, and control thin water films on vertical or inclined substrates. By pinning the film edges like the extremities of a stretched sheet, these grooves prevent collapse and enable the formation of a continuous water layer. This is achieved without chemical coatings or nanoscale surface texturing, offering a fundamentally new and scalable strategy for water control.

Once anchored between the grooves, the water film stretches vertically over more than 100 capillary lengths, stabilised purely by geometric confinement. Its thickness is continuously tunable through the imposed flow rate and groove geometry, and remains stable until it reaches the groove terminus. There, the sudden loss of lateral pinning triggers a clean, localised rupture, initiating a self-sustained cycle of film retraction, puddle inflation, and droplet release. Both film thickness and retraction height follow predictive scaling laws, while the droplet mass also varies systematically with spacing.

This groove-based strategy expands the possibilities for passive water control on vertical surfaces. Beyond simply guiding flow, it enables the formation, stabilisation, and timed rupture of thin films, all using pure water on untreated substrates. These features make it a powerful tool for studying thin-film instabilities under tunable conditions, and open new pathways for robust drainage, condensation management, and programmable droplet release. By combining groove-guided transport with controlled film dynamics, this approach also lays the groundwork for two-dimensional millifluidic systems based on simple, laser-engraved substrates.

While this study establishes a new mechanism for film stabilisation and controlled droplet release, several questions remain. What is the minimal thickness that can be reliably sustained between grooves? Can this principle be scaled up through networks of grooves, or adapted to more complex patterns with non-parallel or staggered terminations? Addressing these questions would not only refine the underlying physical understanding but also expand the practical scope of groove-stabilised films as a platform for passive, programmable fluid control.

\section{Acknowledgment}

The authors would like to thank Valsem Industries SAS, the FNRS CDR project number J.0186.23, the CESAM Research Unit, and the King Abdullah University of Science and Technology (KAUST) for funding this research. The authors also thank Prof. Parmentier for the microscope measurements.

 \end{document}